\documentclass[showpacs,showkeys,preprintnumbers, ]{revtex4}
\usepackage{amssymb}

\usepackage{graphicx}
\usepackage{epsfig}
\usepackage{bm}
\usepackage{fontenc}
\usepackage{graphicx}
\usepackage[dvips]{hyperref}
\usepackage{bm}
\usepackage{amsmath}

\begin{document}

\title{Dirac quasinormal modes of two-dimensional charged dilatonic
black holes}
\author{Ram\'{o}n B\'{e}car}
\email{rbecar@uct.cl}
\affiliation{Departamento de Ciencias Matem\'{a}ticas y F\'{\i}sicas, Universidad Cat\'{o}%
lica de Temuco, Montt 56, Casilla 15-D, Temuco, Chile}
\author{P. A. Gonz\'{a}lez}
\email{pablo.gonzalez@udp.cl}
\affiliation{Facultad de Ingenier\'{\i}a, Universidad Diego Portales, Avenida Ej\'{e}%
rcito Libertador 441, Casilla 298-V, Santiago, Chile.}
\author{Yerko V\'{a}squez.}
\email{yerko.vasquez@ufrontera.cl}
\affiliation{Departamento de F\'{\i}sica, Facultad de Ciencias, Universidad de La Serena,%
\\
Avenida Cisternas 1200, La Serena, Chile.}
\date{\today }

\begin{abstract}
We study charged fermionic perturbations in the background of two-dimensional charged dilatonic black holes, and we present the exact Dirac quasinormal modes. Also, we study the stability of these black holes under charged fermionic perturbations.

\end{abstract}

\maketitle

%\vspace{5cm}

\section{Introduction}

The Einstein-Hilbert action is just a topological
invariant in two spacetime dimensions (Gauss-Bonnet term) and, therefore, it is
necessary to incorporate extra fields to add richness to the two-dimensional gravity models. In this sense, the dilatonic field plays
the role of the extra fields, which naturally arises, for instance, in the
compactifications from higher dimensional models or from string theory. These
theories also have black hole solutions which play an important role in
revealing various aspects about the geometry of spacetime and the quantization of
gravity, and also the physics related to string theory \cite{Witten:1991yr,
Teo:1998kp, McGuigan:1991qp}. On the other hand, two-dimensional low-energy
string theory admits several black hole solutions. Furthermore, technical
simplifications in two dimensions often lead to exact results, and it is
hoped that this might helps to address some  conceptual problems posed
by quantum gravity in higher dimensions. The exact solvability of two-dimensional models of gravity have proven to be a useful tool for
investigations into black hole thermodynamics \cite%
{Lemos:1996bq,Youm:1999xn, Davis:2004xb, Grumiller:2007ju, Quevedo:2009ei,
Belhaj:2013vza}. It is hoped that such investigations provide a deeper
understanding of key issues; including the microscopic origin of black hole
entropy \cite{Myers:1994sg, Sadeghi:2007kn, Hyun:2007ii}, and the end point
of black hole evaporation via thermal radiation \cite{Kim:1999ig,
Vagenas:2001sm, Easson:2002tg}. For an excellent review about dilaton
gravity in two dimensions see \cite{Grumiller:2002nm}. Moreover, there is a growing interest in dilatonic black holes in the last few years,
since it is believed that these black holes can shed some light into the solution of the fundamental problem of the
microscopic origin of the Bekenstein-Hawking entropy. The area-entropy relation $S_{BH} = A/4$ was obtained for a class of five-dimensional extremal black holes in Type II string theory using D-brane techniques \cite{Strominger:1996sh}. In \cite{Teo:1998kp}, the author derived the entropy for the two-dimensional black hole \cite{McGuigan:1991qp} by establishing the U-duality between the two-dimensional black hole and the five-dimensional one \cite{Teo:1998kp}. A similar work was carried out in \cite{LopesCardoso:1998sq} using a different sequence duality transformations, this time in four dimensions, and leading to the same expressions for the entropy for two-dimensional black holes.

The particular motivation of this work is to calculate the quasinormal modes
(QNMs) for charged fermionic field perturbations in the background of two-dimensional charged
dilatonic black holes \cite{McGuigan:1991qp}, and study the stability of these
black holes under charged fermionic perturbations. The QNMs and their quasinormal
frequencies (QNFs) are an important property of black holes and have a long
history, \cite{Regge:1957td, Zerilli:1971wd, Zerilli:1970se,
Kokkotas:1999bd, Nollert:1999ji, Konoplya:2011qq}. 
The QNMs give information about the stability of black holes under matter
fields that evolves perturbatively in the exterior region of them, without
backreacting on the metric. Also, the QNMs determine how fast a thermal
state in the boundary theory will reach thermal equilibrium according to the
AdS/CFT correspondence \cite{Maldacena:1997re}, where the relaxation time of
a thermal state of the boundary thermal theory is proportional to the
inverse of the imaginary part of the QNFs of the dual gravity background 
\cite{Horowitz:1999jd}. In the context of black hole thermodynamics, the
QNMs allow the quantum area spectrum of the black hole horizon to be
studied, as well as the mass and the entropy spectrum. In this regard,
Bekenstein \cite{Bekenstein:1974jk} was the first to propose the idea that
in quantum gravity the area of black hole horizon is quantized, leading to a
discrete spectrum which is evenly spaced. Then, Hod \cite{Hod:1998vk}
conjectured that the asymptotic QNF is related to the quantized black hole
area, by identifying the vibrational frequency with the real part of the
QNFs. However, it is not universal for every black hole background. Then,
Kunstatter \cite{Kunstatter:2002pj} propose that the black hole spectrum can
be obtained by imposing the Bohr-Sommerfeld quantization condition to an
adiabatic invariant quantity involving the energy and the vibrational
frequency. Furthermore, Maggiore \cite{Maggiore:2007nq} argued that in the
large damping limit the identification of the vibrational frequency with the
imaginary part of the QNF could lead to the Bekenstein universal bound.
Then, the consequences of these proposals were studied in several
spacetimes.
Besides, in \cite{Corda:2012tz, Corda:2012dw, Corda:2013nza, Corda:2013paa} the authors
discuss a connection between Hawking radiation and black hole quasinormal
modes, which is important in the route to quantize gravity, because one can
naturally interpret black hole quasinormal modes in terms of quantum levels.  The issue about classical and quantum stability of two dimensional and five dimensional dilatonic black holes was carried out, for instance in \cite{Kim:1994nq, Nojiri:1998yg, Frolov:2000jh, Becar:2007hu, Becar:2010zz, LopezOrtega:2009zx}, and it was shown that the  absorption cross section vanishes at the low and high frequency limit for these black holes \cite{Becar:2014aka}. Also, the QNMs for charged dilatonic black holes were studied in \cite{Li:2001ct} via semi-analytic and numerical methods.

This paper is organized as follows. In Sec. \ref{FP}, we study charged fermionic
perturbations in the background of two-dimensional charged dilatonic black
holes, and we calculate the exact QNMs.
% of charged fermionic perturbations for two-dimensional charged dilatonic black hole. 
Finally, remarks are presented in Sec. %
\ref{remarks}.

\section{Charged fermionic perturbations of two-dimensional charged dilatonic black holes}

\label{FP} 
Maxwell-gravity coupled to a
dilatonic field ($\phi$) can be described by the effective action \cite{McGuigan:1991qp} 
\begin{equation}
S=\frac{1}{2\pi }\int d^{2}x\sqrt{-g}e^{-2\phi }\left( R-4(\nabla \phi
)^{2}-\lambda -\frac{1}{4}F_{\mu \nu }F^{\mu \nu }\right)~,  \label{accion}
\end{equation}%
where $R$ is the Ricci scalar, $\lambda$ is the effective central charge and $F_{\mu\nu}$ is the electromagnetic strength tensor. 
The equations of motion for the metric, gauge and dilaton field are given by 
\begin{eqnarray}
\nonumber R_{\mu \nu }-2\nabla _{\mu }\nabla _{\nu }\phi -\frac{1}{2}F_{\mu \sigma
}F_{\nu }^{\sigma } &=&0~,  \label{betag} \\
\nonumber \nabla _{\nu }\left( e^{-2\phi }F^{\mu \nu }\right)  &=&0\text{ }, \\
R-4\nabla _{\mu }\nabla ^{\mu }\phi +4\nabla _{\mu }\phi \nabla ^{\mu }\phi
-\lambda -\frac{1}{4}F_{\mu \nu }F^{\mu \nu } &=&0~.  \label{betaphi}
\end{eqnarray}%
The general static metric describing charged black holes in this theory can
be written as 
\begin{equation}
ds^{2}=-f(r)dt ^{2}+\frac{dr^{2}}{f(r)},  \label{metrica1}
\end{equation}%
where $f(r)=1-2me^{-Qr}+q^{2}e^{-2Qr}$, $\phi =\phi _{0}-\frac{Q}{2}r$, and $%
F_{tr}=\sqrt{2}Qqe^{-Qr}$. The condition of
asymptotic flatness for the spacetime requires $\lambda=-Q^2$. The free parameter $m$ is proportional to the black hole mass, and $q$ is proportional to the black hole electric charge. The horizons of the black hole are located at
\begin{equation}
r_{\pm }=\frac{1}{Q}\ln \left( m\pm \sqrt{m^{2}-q^{2}}\right) \text{ },
\end{equation}
therefore, the condition $m^{2}-q^{2}\geqslant 0$ must be satisfied
in order to have an event horizon at $r_{+}$. The change of coordinate $y=e^{-Qr}$, yields $f(y)=1-2my+q^{2}y^{2}$ where the spatial infinity is now located at $y=0$. This solution represents a
well-known string-theoretic black hole \cite%
{Teo:1998kp,McGuigan:1991qp,Witten:1991yr}. The charged fermionic
perturbations in the background of two-dimensional dilatonic black holes are governed by the Dirac equation in curved space
\begin{equation}
\left( \gamma ^{\mu }\left( \nabla _{\mu }-iq^{\prime }A_{\mu }\right)
+m^{\prime }\right) \psi =0~,  \label{DE}
\end{equation}%
where $A_{\mu}$ is the electromagnetic potential, $q^{\prime }$ and $m^{\prime }$ are the electric charge and the mass of the fermionic field $\psi$ respectively, 
and the covariant derivative is defined as 
\begin{equation}
\nabla _{\mu }=\partial _{\mu }+\frac{1}{2}\omega _{\text{ \ \ \ }\mu
}^{ab}J_{ab}~,
\end{equation}%
where $J_{ab}=\frac{1}{4}\left[ \gamma _{a},\gamma _{b}\right]$ are the generators of the Lorentz group. The gamma matrices in curved spacetime $\gamma ^{\mu }$ are defined by 
%\begin{equation}
$\gamma ^{\mu }=e_{\text{ \ }a}^{\mu }\gamma ^{a}$,
%\end{equation}%
where $\gamma ^{a}$ are the gamma matrices in flat spacetime. In order to
solve the Dirac equation, we use the diagonal vielbein 
\begin{equation}
e^{0}=\sqrt{f\left( r\right) }dt~,\text{ \ }e^{1}=\frac{1}{\sqrt{f\left(
r\right) }}dr~.
\end{equation}%
From the null torsion condition 
%\begin{equation}
$de^{a}+\omega _{\text{ \ }b}^{a}\wedge e^{b}=0$, 
% \label{NT}
%\end{equation}%
we obtain the spin connection 
\begin{equation}
\omega ^{01}=\frac{f^{\prime }\left( r\right) }{2\sqrt{f\left( r\right) }}%
e^{0}.
\end{equation}%
Now, by using the following representation of the gamma matrices 
\begin{equation}
\gamma ^{0}=i\sigma ^{2}~,\text{ \ }\gamma ^{1}=\sigma ^{1}~,
\end{equation}%
where $\sigma ^{i}$ are the Pauli matrices, 
along with the following ansatz for the fermionic field 
\begin{equation}
\psi =\frac{1}{f\left( r\right) ^{1/4}}e^{-i\omega t}\left( 
\begin{array}{c}
\psi _{1} \\ 
\psi _{2}%
\end{array}%
\right) ~,
\end{equation}%
we obtain the following equations 
\begin{eqnarray}
\sqrt{f}\partial _{r}\psi _{1}+\frac{i\omega }{\sqrt{f}}\psi _{1}+\frac{%
\sqrt{2}iqq^{\prime }}{\sqrt{f}}e^{-Qr}\psi _{1}+m^{\prime }\psi _{2} &=&0~  \notag
\label{system} \\
\sqrt{f}\partial _{r}\psi _{2}-\frac{i\omega }{\sqrt{f}}\psi _{2}-\frac{%
\sqrt{2}iqq^{\prime }}{\sqrt{f}}e^{-Qr}\psi _{2}+m^{\prime }\psi _{1} &=&0~.
\end{eqnarray}
Decoupling the above system of equations, we obtain the following
equation for $\psi _{1}$ 
\begin{gather}
\nonumber 2f\left( r\right) ^{2}\psi _{1}^{\prime \prime }(r)+f\left( r\right)
f^{\prime }\left( r\right) \psi _{1}^{\prime }(r)+e^{-2Qr}(4q^{2}q^{\prime
2}+4\sqrt{2}e^{Qr}qq^{\prime }\omega +2e^{2Qr}\omega ^{2}   \\
-2e^{Qr}(m^{\prime 2}e^{Qr}+\sqrt{2}iqq^{\prime }Q)f(r)-ie^{Qr}\left( \sqrt{2}%
qq^{\prime }+e^{Qr}\omega \right) f^{\prime }(r))\psi _{1}(r)=0~.
\label{radial}
\end{gather}
Now, making the change of variables $y=e^{-Qr}$, and after some algebraic manipulation, the equation (\ref%
{radial}) becomes 
\begin{equation}
\psi _{1}^{\prime \prime }\left( y\right) +\left( \frac{1}{y}+\frac{1/2}{%
y-y_{+}}+\frac{1/2}{y-y_{-}}\right) \psi _{1}^{\prime }\left( y\right)
+\left( \frac{A_{1}}{y}+\frac{A_{2}}{y-y_{+}}+\frac{A_{3}}{y-y_{-}}\right) 
\frac{1}{y\left( y-y_{+}\right) \left( y-y_{-}\right) }\psi _{1}\left(
y\right) =0~.  \label{diff}
\end{equation}
Where $y_{\pm }$ are the roots of $f(y)=1-2my+q^{2}y^{2}$, and are
given explicitly by
\begin{equation}
y_{\pm }=\frac{m\mp \sqrt{m^{2}-q^{2}}}{q^{2}}\text{ },
\end{equation}
and the constants $A_{1}$, $A_{2}$ and $A_{3}$ are defined by the following
expressions
\begin{eqnarray}
A_{1} &=&\frac{1}{q^{2}Q^{2}}\left( \omega ^{2}-m^{\prime 2}\right) \text{ }, \\
A_{2} &=&y_{+}\left( y_{+}-y_{-}\right) \left( \frac{1}{16}-\left( \frac{1}{4%
}-\frac{i\omega }{q^{2}Qy_{+}\left( y_{+}-y_{-}\right) }-\frac{\sqrt{2}%
iq^{\prime }}{qQ\left( y_{+}-y_{-}\right) }\right) ^{2}\right) \text{ }, \\
A_{3} &=&-y_{-}\left( y_{+}-y_{-}\right) \left( \frac{1}{16}-\left( \frac{1}{%
4}+\frac{i\omega }{q^{2}Qy_{-}\left( y_{+}-y_{-}\right) }+\frac{\sqrt{2}%
iq^{\prime }}{qQ\left( y_{+}-y_{-}\right) }\right) ^{2}\right) \text{ }.
\end{eqnarray}
Furthermore, performing another change of variable $z=\frac{y_{+}-y}{y_{+}}$, equation (%
\ref{diff}) reduces to
\begin{gather}
\nonumber \psi _{1}^{\prime \prime }\left( z\right) +\left( \frac{1}{z-1}+\frac{1/2}{z}%
+\frac{1/2}{z-1+\frac{y_{-}}{y_{+}}}\right) \psi _{1}^{\prime }\left(
z\right) +  \\
\frac{1}{y_{+}^{2}}\left( \frac{A_{1}}{z-1}+\frac{A_{2}}{z}+\frac{A_{3}}{z-1+%
\frac{y_{-}}{y_{+}}}\right) \frac{1}{z\left( z-1\right) \left( z-1+\frac{%
y_{-}}{y_{+}}\right) }\psi _{1}\left( z\right) =0~.  
 \label{ecuacion}
\end{gather}
We note that the above equation corresponds to the Riemann's differential equation,
whose general form is \cite{M. Abramowitz} 
\begin{eqnarray}
\nonumber &&\frac{d^{2}w}{dz^{2}}+\left( \frac{1-\alpha -\alpha ^{\prime }}{z-r}+\frac{%
1-\beta -\beta ^{\prime }}{z-s}+\frac{1-\gamma -\gamma ^{\prime }}{z-t}%
\right) \frac{dw}{dz}+  \label{eqn} \\
&&\left( \frac{\alpha \alpha ^{\prime }\left( r-s\right) \left( r-t\right) }{%
z-r}+\frac{\beta \beta ^{\prime }\left( s-t\right) \left( s-r\right) }{z-s}+%
\frac{\gamma \gamma ^{\prime }\left( t-r\right) \left( t-s\right) }{z-t}%
\right) \frac{w}{\left( z-r\right) \left( z-s\right) \left( z-t\right) }=0~,
\end{eqnarray}%
where $r,s,t$ are the singular points, and the exponents $\alpha ,\alpha
^{\prime },\beta ,\beta ^{\prime },\gamma ,\gamma ^{\prime }$ are subject to
the condition 
\begin{equation}
\alpha +\alpha ^{\prime }+\beta +\beta ^{\prime }+\gamma +\gamma ^{\prime
}=1~.
\end{equation}%
The complete solution of (\ref{eqn}) is denoted by the symbol 
\begin{equation}
w=P\left\{ 
\begin{array}{cccc}
r & s & t &  \\ 
\alpha & \beta & \gamma & z \\ 
\alpha ^{\prime } & \beta ^{\prime } & \gamma ^{\prime } & 
\end{array}%
\right\} ~,
\end{equation}%
where the $P$ symbol denotes the Riemann's $P$ function, which can be
reduced to the hypergeometric function through 
\begin{equation}
w=\left( \frac{z-r}{z-s}\right) ^{\alpha }\left( \frac{z-t}{z-s}\right)
^{\gamma }P\left\{ 
\begin{array}{cccc}
0 & \infty & 1 &  \\ 
0 & \alpha +\beta +\gamma & 0 & \frac{\left( z-r\right) \left( t-s\right) }{%
\left( z-s\right) \left( t-r\right) } \\ 
\alpha ^{\prime }-\alpha & \alpha +\beta ^{\prime }+\gamma & \gamma ^{\prime
}-\gamma & 
\end{array}%
\right\} ~,
\end{equation}%
where the $P$ function is now the Gauss' hypergeometric function. So, considering equations (\ref{ecuacion}) and (\ref{eqn}) we can identify
the regular singular points $r, s$ and $t$ as 
\begin{equation}
r=0~,\text{ \ }s=1-\frac{y_{-}}{y_{+}}~,\text{ \ }t=1~.
\end{equation}%
Therefore, the exponents are given by 
\begin{eqnarray}
\alpha &=&-\frac{i\omega }{q^{2}Qy_{+}\left( y_{-}-y_{+}\right) }~-\frac{%
\sqrt{2}iq^{\prime }}{qQ\left( y_{-}-y_{+}\right) },\ \ \alpha ^{\prime }=%
\frac{1}{2}+\frac{i\omega }{q^{2}Qy_{+}\left( y_{-}-y_{+}\right) }~+\frac{%
\sqrt{2}iq^{\prime }}{qQ\left( y_{-}-y_{+}\right) }~, \\
\beta &=&\frac{i\omega }{q^{2}Qy_{-}\left( y_{-}-y_{+}\right) }~+\frac{\sqrt{%
2}iq^{\prime }}{qQ\left( y_{-}-y_{+}\right) },\ \ \ \ \ \beta ^{\prime }=%
\frac{1}{2}-\frac{i\omega }{q^{2}Qy_{-}\left( y_{-}-y_{+}\right) }~-\frac{%
\sqrt{2}iq^{\prime }}{qQ\left( y_{-}-y_{+}\right) }~, \\
\gamma &=&\frac{\sqrt{m^{\prime 2}-\omega ^{2}}}{Q}~,\ \ \ \ \ \ \ \ \ \ \ \ \ \ \
\ \gamma ^{\prime }=-\frac{\sqrt{m^{\prime 2}-\omega ^{2}}}{Q}~,
\end{eqnarray}%
and the solution to equation (\ref{diff}) can be written as 
\begin{eqnarray}
\psi _{1}\left( z\right) &=&C_{1}\left( \frac{z}{z-1+\frac{y_{-}}{y_{+}}}%
\right) ^{\alpha }\left( \frac{z-1}{z-1+\frac{y_{-}}{y_{+}}}\right) ^{\gamma
}{_{2}}F{_{1}}\left( a,b,c,\frac{\frac{y_{-}}{y_{+}}z}{z-1+\frac{y_{-}}{y_{+}%
})}\right) +  \notag \\
&&C_{2}\left( \frac{z}{z-1+\frac{y_{-}}{y_{+}}}\right) ^{\alpha ^{\prime
}}\left( \frac{z-1}{z-1+\frac{y_{-}}{y_{+}}}\right) ^{\gamma }{_{2}}F{_{1}}%
\left( a-c+1,b-c+1,2-c,\frac{\frac{y_{-}}{y_{+}}z}{z-1+\frac{y_{-}}{y_{+}})}%
\right) ~,
\end{eqnarray}%
where we have defined the constants $a,b$ and $c$ as 
\begin{eqnarray}
a &=&\alpha +\beta +\gamma ~,  \notag \\
b &=&\alpha +\beta ^{\prime }+\gamma ~,  \notag \\
c &=&1+\alpha -\alpha ^{\prime }~.
\end{eqnarray}%
In the near horizon limit, the above expression behaves as 
\begin{equation}
\psi _{1}\left( z\rightarrow 0\right) =\widehat{C}_{1}z^{\alpha }+\widehat{C}%
_{2}z^{\alpha ^{\prime }}~,
\end{equation}%
where $\widehat{C}_{1}$ and $\widehat{C}_{2}$ are constants.
Now, we impose as boundary condition that classically nothing can escape
from the event horizon, therefore we must take $C_{2}=0$ in order to have
only ingoing waves at the horizon. So, the solution simplifies to 
\begin{equation}
\psi _{1}\left( z\right) =C_{1}\left( \frac{z}{z-1+\frac{y_{-}}{y_{+}}}%
\right) ^{\alpha }\left( \frac{z-1}{z-1+\frac{y_{-}}{y_{+}}}\right) ^{\gamma
}{_{2}}F{_{1}}\left( a,b,c,\frac{\frac{y_{-}}{y_{+}}z}{z-1+\frac{y_{-}}{y_{+}%
})}\right) ~.
\end{equation}%
Now, we implement boundary conditions at the spatial infinity $z\rightarrow 1$.
In order to do this, we employ the Kummer's relations \cite{M. Abramowitz}
\begin{gather}
{_{2}}F_{1}\left( a,b,c;z\right) =\frac{\Gamma \left( c\right) \Gamma \left(
c-a-b\right) }{\Gamma \left( c-a\right) \Gamma \left( c-b\right) }{_{2}}%
F_{1}\left( a,b,a+b-c;1-z\right) +  \notag \\
\left( 1-z\right) ^{c-a-b}\frac{\Gamma \left( c\right) \Gamma \left(
a+b-c\right) }{\Gamma \left( a\right) \Gamma \left( b\right) }{_{2}}%
F_{1}\left( c-a,c-b,c-a-b+1;1-z\right) ~,
\label{relationkummer}
\end{gather}%

which allow us to write the solution as 
\begin{eqnarray}
\psi _{1}\left( z\right) &=&C_{1}\left( \frac{z}{z-1+\frac{y_{-}}{y_{+}}}%
\right) ^{\alpha }\left( \frac{z-1}{z-1+\frac{y_{-}}{y_{+}}}\right) ^{\gamma
}\frac{\Gamma \left( c\right) \Gamma \left( c-a-b\right) }{\Gamma \left(
c-a\right) \Gamma \left( c-b\right) }{_{2}}F{_{1}}\left( a,b,a+b-c,1-\frac{%
\frac{y_{-}}{y_{+}}z}{z-1+\frac{y_{-}}{y_{+}})}\right) +  \notag \\
&&C_{1}\left( 1-\frac{y_{-}}{y_{+}}\right) ^{\gamma ^{\prime }-\gamma
}\left( \frac{z}{z-1+\frac{y_{-}}{y_{+}}}\right) ^{\alpha }\left( \frac{z-1}{%
z-1+\frac{y_{-}}{y_{+}}}\right) ^{\gamma ^{\prime }}\frac{\Gamma \left(
c\right) \Gamma \left( a+b-c\right) }{\Gamma \left( a\right) \Gamma \left(
b\right) }\times  \notag \\
&&{_{2}}F{_{1}}\left( c-a,c-b,c-a-b+1,1-\frac{\frac{y_{-}}{y_{+}}z}{z-1+%
\frac{y_{-}}{y_{+}})}\right) ~.
\end{eqnarray}%
In the limit $z\rightarrow 1$, the above expression becomes 
\begin{equation}
\psi _{1}\left( z\rightarrow 1\right) =\widetilde{C}_{1}\left( 1-z\right)
^{\gamma }\frac{\Gamma \left( c\right) \Gamma \left( c-a-b\right) }{\Gamma
\left( c-a\right) \Gamma \left( c-b\right) }+\widetilde{C}_{1}\left( 1-\frac{%
y_{-}}{y_{+}}\right) ^{\gamma ^{\prime }-\gamma }\left( 1-z\right) ^{\gamma
^{\prime }}\frac{\Gamma \left( c\right) \Gamma \left( a+b-c\right) }{\Gamma
\left( a\right) \Gamma \left( b\right) }~,  \label{infinity}
\end{equation}
where $\widetilde{C}_{1}$ is a constant.
So, in order to have only outgoing waves at the spatial infinity $z=1$,
we must impose $c-a=-n$ or $c-b=-n$. These conditions yield the following
sets of quasinormal frequencies for two-dimensional charged dilatonic
black holes
\begin{eqnarray}
\nonumber \omega_1  &=&-\frac{\sqrt{2}mqq^{\prime }}{q^{2}}-i\frac{mQ\sqrt{m^{2}-q^{2}}}{%
2q^{2}}\left( 1+2n\right) + \\
&&i\frac{\sqrt{m^{2}-q^{2}}}{2q^{2}}\sqrt{Q^{2}\left( m^{2}-q^{2}\right)
\left( 1+2n\right) ^{2}+4q^{2}\left( m^{\prime 2}-2q^{\prime 2}\right) -4\sqrt{2}%
iqq^{\prime }Q\sqrt{m^{2}-q^{2}}\left( 1+2n\right) }~,
\end{eqnarray}
and
\begin{equation}\label{w2}
\omega_2 =-i\frac{nQ}{2}+i\frac{m^{\prime 2}}{2nQ}\text{ },
\end{equation}
where $n=0,1,2,...$ for the first set and $n=1,2,...$ for the second set of frequencies. Depending on the value of the fermionic field mass $m^{\prime}$, some frequencies can have
positive imaginary part, and therefore the black holes can be unstable. In order to separate the real part from the imaginary part for the first set of frequencies, we define
\begin{equation}
Z=Q^{2}\left( m^{2}-q^{2}\right)
\left( 1+2n\right) ^{2}+4q^{2}\left( m^{\prime 2}-2q^{\prime 2}\right) -4\sqrt{2}%
iqq^{\prime }Q\sqrt{m^{2}-q^{2}}\left( 1+2n\right)~, 
\end{equation} 
therefore, the first set of frequencies can be written as
\begin{equation}\label{omega1}
 \omega_1  =-\frac{\sqrt{2}mqq^{\prime }}{q^{2}}+ sgn(qq^{\prime})\frac{\sqrt{m^{2}-q^{2}}}{2q^{2}} \sqrt{\frac{\left \vert Z \right\vert-Re(Z)}{2}}-i\frac{\sqrt{m^{2}-q^{2}}}{%
2q^{2}}\left(mQ\left( 1+2n\right) - \sqrt{\frac{\left \vert Z \right\vert+Re(Z)}{2}}\right)~,
\end{equation}
where $sgn(qq^{\prime})$ gives the sign of $qq^{\prime}$. Therefore, the imaginary part of the first set of frequencies is negative for
\begin{equation}\label{conditionm}
m^{\prime 2}<\frac{m^2Q^2(1+2n)^2+8q^2q^{\prime 2}}{4m^2}~.
\end{equation}
So, if the above inequality is satisfied, the two-dimensional charged dilatonic black holes are stable under charged fermionic perturbations. Note that in the case $q^{\prime }\neq 0$ the QNFs acquire a real part. Analogously, for $m^{\prime 2}<n^2Q^2$, the imaginary part of the second set of frequencies is negative, which guaranties the stability. Remarkably, this second set is the same as the QNFs for  two-dimensional uncharged dilatonic black holes \cite{Becar:2007hu, LopezOrtega:2009zx}.     
In a similar way, the QNFs associated to $\psi _{2}$ can
be obtained, note that $\psi _{2}$ satisfies a similar equation that $\psi
_{1}$ but making the changes $q^{\prime} \rightarrow - q^{\prime}$ and $\omega
\rightarrow -\omega $.
For $\psi _{2}$, and following an analogous procedure, we find two additional sets of quasinormal frequencies, one of them is the same quasinormal frequencies given in eq. (\ref{omega1}) and the other set is given by
\begin{equation}\label{w2}
\omega_3 =-i\frac{ \left (n+1 \right ) Q}{2}+i\frac{m^{\prime 2}}{2\left (n+1 \right ) Q}\text{ },
\end{equation}

where $n=0,1,2, ...$. In this case, the imaginary part is negative when the fermionic field mass satisfies the inequality $m^{\prime 2}<\left (n+1 \right )^2Q^2$.
\section{Final remarks}

\label{remarks}

In this work we have computed analytically the QNMs of charged fermionic
perturbations for two-dimensional charged dilatonic black holes and we analyzed the stability of these black holes under fermionic perturbations. The fermionic fields are solutions of the Dirac equation, which can be reduced to the Riemann's differential equation, as in \cite{Catalan:2014ama, Gonzalez:2014voa}, and we have shown that there are two set of quasinormal frequencies for each component of the fermionic field, where the second set (\ref{w2}), is the same that the uncharged case \cite{Becar:2007hu, LopezOrtega:2009zx}. On the other hand, the first set of quasinormal frequencies, are given by a real part when $q^{\prime}\neq 0$, and an imaginary part that can be negative if the square fermionic field mass satisfies (\ref{conditionm}).

%%%%%%%%%%%%%%%%%%%%%%%%%%%%%%%%%%%%%%%%%%%%%%%%%%%%%%%%%%%%%%%%%

%%%%%%%%%%%%%%%%%%%%%%%%%%%%%%%%%%%%%%%%%%%%%%%%%%%%%%%%%%%%%%%%%%%%%%%%%%%%%%%%%%%%
%%%%%%%%%%%%%%%%%%%%%%%%%%%%%%%%%%%%%%%%%%%%%%%%%%%%%%%%%%%%%%%%%%%%%%%%%%%%%%%%%%%%

\section*{Acknowledgments}

This work was funded by Comisi{\'o}n Nacional de Investigaci{\'o}n Cient{\'i}%
fica y Tecnol{\'o}gica through FONDECYT Grant 11121148 (Y.V.).

\end{document}